# Fundamental Parameters of Some New Discovered Open star Clusters

El-Bendary R., Tadross A. L.

*National Research Institute of Astronomy and Geophysics, Cairo, Egypt*

## ABSTRACT

We present here the fundamental properties of some newly discovered open star clusters (Teutsch 144, Alessi 53, Riddle 4 and Juchrt 12) using the *JHK* Near-IR photometry (*2MASS survey*) of Cutri et al. (2003). These clusters have been selected from Kronberger et al. (2006) who presented some new discovered stellar groups on the basis of *2MASS* photometry and the *DSS* visual images. The astrometry and photometric parameters are determined using the stellar density distributions and color-magnitude diagrams fittings. Center, radius, membership, distances, reddening, age, luminosity function, mass function, total mass, and the dynamical relaxation time have been estimated for the first time. This paper is a part of Reda's PhD project.

# 1. Introduction

Nowadays, stellar structure and evolution are among the most interesting topics in modern astronomy. Stellar groups (associations, open and globular clusters) occupy a predominant place in this respect. Open star clusters, which are formed along the gas and dust of the Galactic plane contain from tens to a few thousands stars distributed in an approximately spherical structure of up to a few parsecs in radius. This makes them potentially short-lived stellar systems. Consequently, most of them are evaporated completely in less than 1 Gyr (Bonatto et al. 2004).

Kronberger et al. (2006) presented a list of some stellar groups which are not sure if they are open clusters or not. In Reda's PhD plan, more than 20 stellar groups have been selected for his project, while 4 groups only (Teutsch 144, Alessi 53, Riddle 4 and Juchrt 12) are selected for the present work as a first part of his work. Fig. 1 represents the images of the clusters as taken from Digitized Sky Surveys (DSS).

In sect. 2, the data extraction has been presented; the astrometry and photometric analyses with CMDs are presented in sect. 3 and 4 respectively. In sect. 5, the luminosity and mass functions are presented. Dynamical state is introduced in sect. 6. Finally, the conclusion has been devoted in sec. 7.

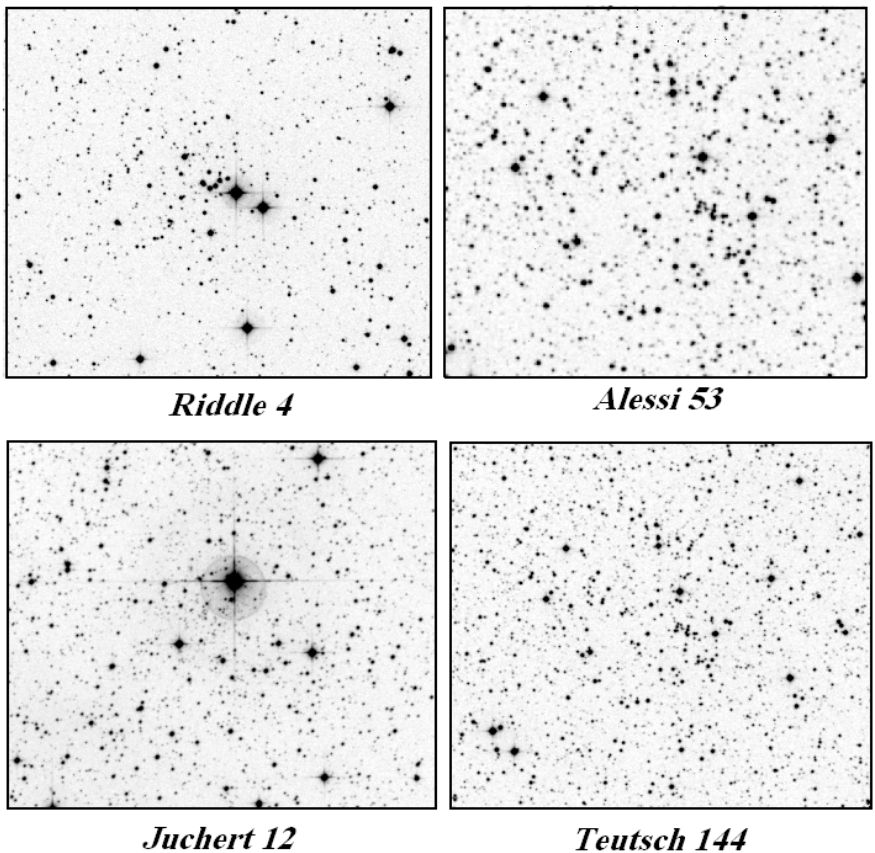

**Fig. 1: The images of the clusters as taken from Digitized Sky Surveys. North is up and east on the left.**

## 2. Data Extraction

We investigate the nature and structure of each cluster using J, H, and K photometry obtained from 2MASS data release which is available at:

http://www.ipac.caltech.edu/2mass/releases/allsky/.

It is uniformly scanning the entire sky in three near-IR bands J (1.25$\mu$m), H (1.65$\mu$m) and $K_s$ (2.17$\mu$m) with two highly automated 1.3m telescopes equipped with three channel camera, each channel consisting of a 256 x 256 array of the HgCdTe detectors.

The only information known about these clusters is the coordinates, which are listed here in Table 1.

Table 1: Clusters' coordinates.

| Cluster | $\alpha$ | $\delta$ | l | b |
|---|---|---|---|---|
| **Riddle 4** | 2:7:22.7 | +60:15:25 | 132.222043 | -1.237864 |
| **Alessi 53** | 6:29:24.5 | +9:10:39 | 202.261324 | -0.664421 |
| **Juchert 12** | 7:20:56.7 | -22:52:00 | 236.5614 | -4.1232 |
| **Teutsch 144** | 21:21:43.9 | +50:36:36 | 92.73462 | +0.459096 |

## 3. The Astrometry of the clusters

### 3.1. Centers determination

There are several ways to define the cluster center; *theoretically*, it can be defined as the center of mass or the location of the deepest part of the gravitational potential.

*Observationally*, it defined as the location of maximum stellar density (the number of stars per unit area in the direction of the cluster).The cluster center is found by fitting Gaussian to the profiles of star counts in right ascension and declination, cf. Tadross 2004, and 2005.

### 3.2. Diameters determination

Firstly, in order to determine the cluster radii (real border) the radial surface density of the stars must be achieved. The cluster border is defined as the point at which covers the cluster area and reach enough stability in the background density, i.e., the difference between the observed density profile and the background one is almost equal zero (cf. Tadross 2004). Bonatto et al. (2005) demonstrated that the determination of a cluster radius is made possible by the spatial coverage and uniformity of 2MASS photometry which allows one to obtain reliable data on the projected distribution of stars for large extensions around the clusters' center.

In this context, the cluster area is divided into a number of concentric circles out to the preliminary radius of 20 arcmin. The stellar density (ρ) in each zone has been calculated as N/A, where N is the number of stars at this zone, and A is the area of that zone. Although the spatial shape of the clusters may not be perfectly spherical, the fitting of King (1962) can be applied to derive the clusters radii.

4. **CMDs photometric analysis**

The fundamental parameters for these clusters (age, reddening, and distance) can be determined by fitting the solar theoretical isochrones of Bonatto et al. (2004) to the CMD of each cluster. Several fittings have been applied to our clusters using different age's isochrones. Color and magnitude filters have been applied to the $J \sim (J-H)$ sequences where stars located away from the main sequences are excluded (Bonatto et al. 2005), see Tadross 2008 and references therein. Once the best fit has obtained, we can get the age, distance, and color excess $E_{J-H}$. The optical color excess $E_{B-V}$ can be also estimated; applying Dutra et al. (2002)'s relations. The radius and CMD of each cluster are shown in Figs. 2, a, b, c & d, respectively.

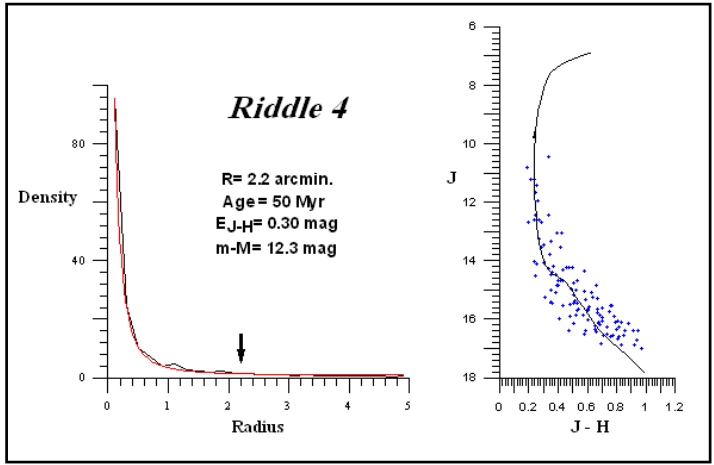

**Fig. 2a: The radius and CMD of Riddle 4.**

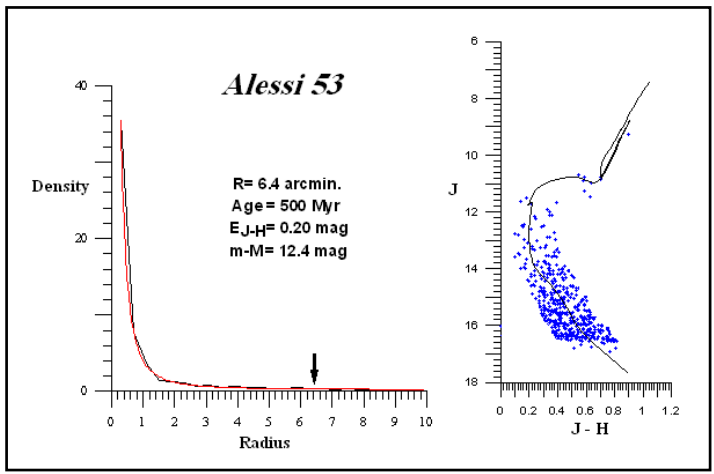

**Fig. 2b: The radius and CMD of Alessi 53.**

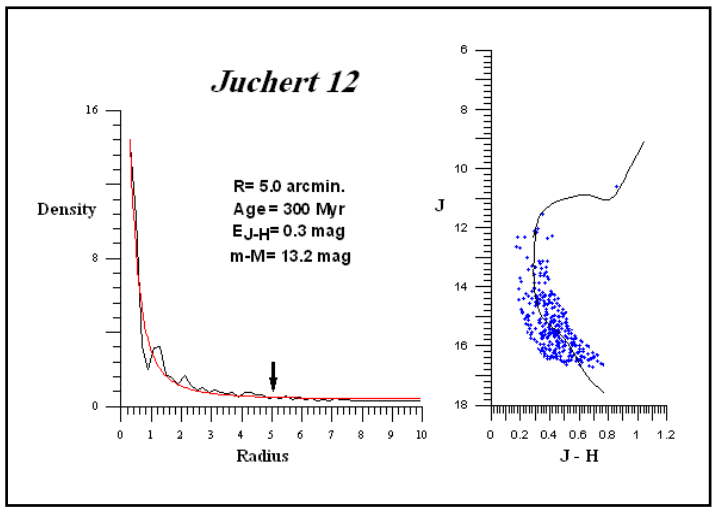

**Fig. 2c: The radius and CMD of Juchert 12.**

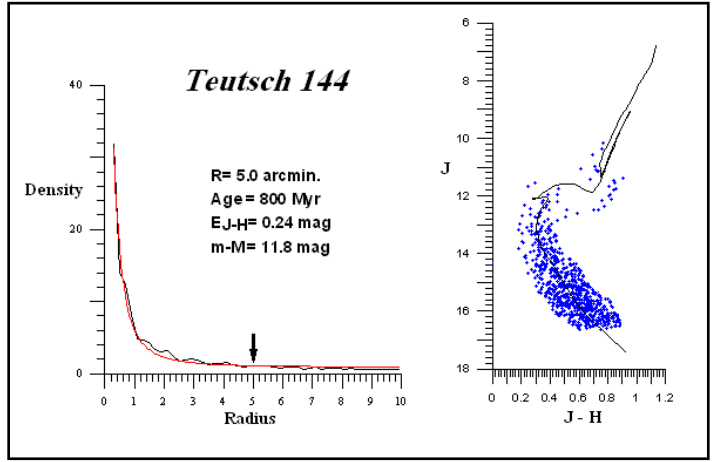

**Fig. 2d: The radius and CMD of Teutsch 144.**

Once the cluster's distance is estimated, distance from the sun, distance from the galactic center, Rgc, and the projected distances on the galactic plane from the Sun, X, Y, and the distance from galactic plane, Z can be determined. Table 2 presents the clusters' properties which are estimated in the present study.

**Table 2: The clusters' properties**

| Cluster | R | Age | $E_{B-V}$ | m-M | Dist. | X | Y | Z | $R_{gc}$ |
|---|---|---|---|---|---|---|---|---|---|
| | arcmin | M yr | mag. | mag. | Pc. | Pc. | Pc. | Pc. | Kpc |
| **Riddle 4** | 2.2 | 50 | 0.91 | 12.3 | 1992 | 1337 | 1473 | -43 | 9.9 |
| **Alessi 53** | 6.4 | 500 | 0.61 | 12.4 | 2360 | 2184 | -894 | -27 | 10.7 |
| **Juchert 12** | 5 | 300 | 0.91 | 13.2 | 3016 | 1657 | -2509 | -217 | 10.5 |
| **Teutsch 144** | 5 | 800 | 0.73 | 11.8 | 1704 | 81 | 1698 | 14 | 8.7 |

Columns from left to right represent the cluster's name; radius; age ; reddening; distance modulus; distances from the sun in parsecs; the projected distances on the galactic plane from the Sun, X, Y, Z in parsecs; and the distance from the galactic center in kilo parsecs, respectively.

## 5. Luminosity and Mass Functions

In order to estimate the luminosity function we can count the observed stars in terms of absolute magnitude after applying the distance modulus, which derived above for these clusters.

The magnitude bin intervals are selected to include a reasonable number of stars in each bin and for the best possible statistics of the luminosity and mass functions *(LF & MF)*. From *LF*, we can infer that the massive bright stars seem to be centrally concentrated more than the low masses and fainter ones (Montgomery et al. 1993).

*LF*s and *MF*s of our clusters are constructed as shown in Figs. 3, a, b, c & d, respectively.

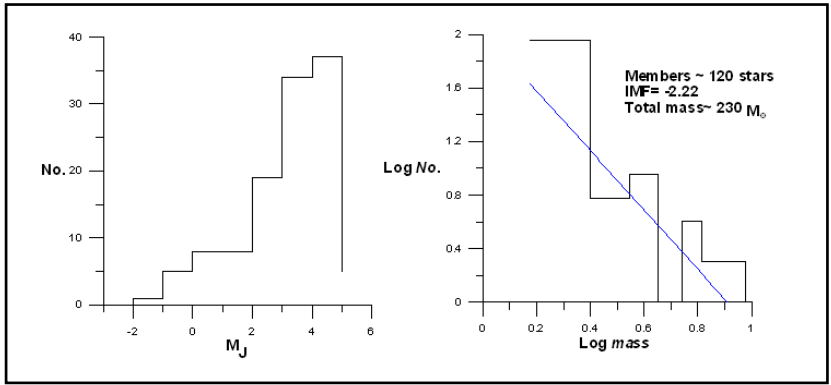
**Fig. 3a: LF and MF for Riddle 4.**

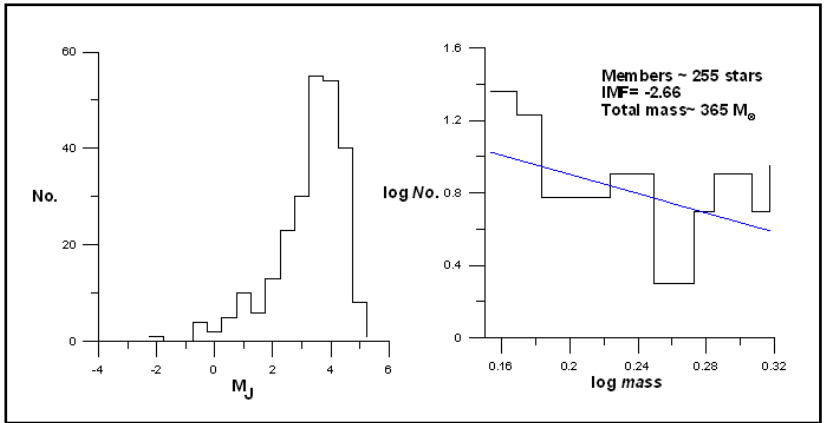
**Fig. 3b: LF and MF for Alessi 53.**

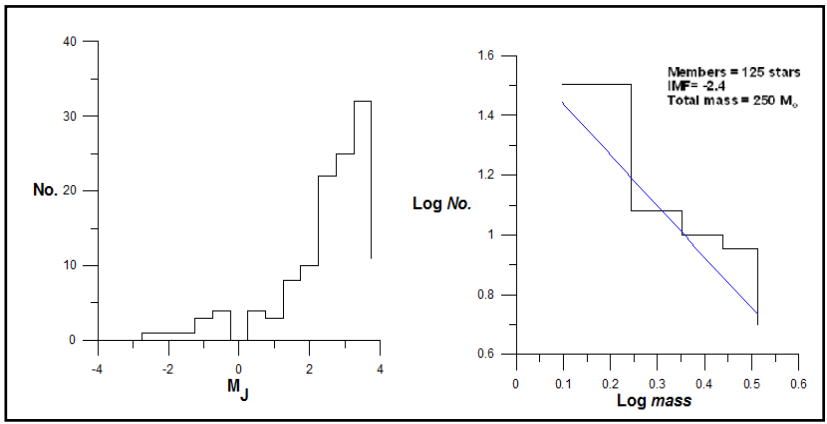
**Fig. 3c: LF and MF for Juchert 12.**

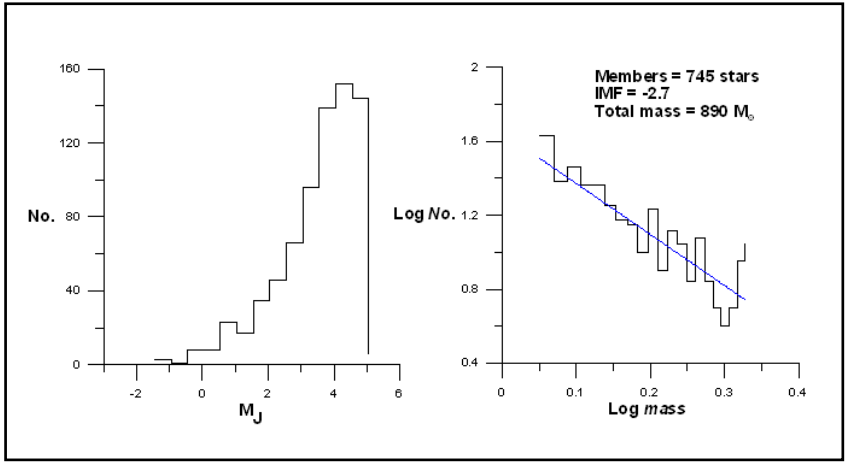

**Fig. 3d: LF and MF for Teutsch 144.**

The stellar initial mass function (IMF) is defined as the density of stars per unit mass bin, and given by the next equation:

$$\psi(M) = \frac{dN}{dM}$$

where dN/dM is the number of stars per unit mass as a function of mass. The IMF for massive stars ($> 1$ M$_\odot$) has been studied and well established by Salpeter (1955).

The *MF* for each cluster has been estimated from a polynomial equation of forth degrees developed from the isochrones data of solar metallicity and the age of each cluster individually. On the other hand, the summation of multiplying the number of stars in each bin by the mean mass of that bin yields a total mass of each cluster. See Table 3.

## 6. Dynamical state

The relaxation time $T_R$ of a cluster is defined as the time, in which the cluster needs from the very beginning to build itself and to reach stability against the contraction and destruction forces; e.g., gas pressure, turbulence, rotation, and the magnetic field (cf. Tadross 2005). $T_R$ is depending mainly on the number of members and the mean radius of the cluster and given by the equation:

$$T_R = \frac{8.9 \times 10^5 N^{1/2} R_h^{3/2}}{<m>^{1/2} \log(0.4N)}$$

Where N is the number of cluster members, $R_h$ is the radius containing half of the cluster mass in parsecs and <m> is the average mass of the cluster stars in solar unit (Spitzer & Hart 1971). By using the above equation we can estimate the dynamical relaxation time of each cluster, as shown in Table 3.

**Table 3: The main parameters of LF & MF of the clusters**

| Cluster | IMF | Members | T. M. $M_\odot$ | $T_R$ M yr |
|---|---|---|---|---|
| Riddle 4 | -2.22 | 120 | 230 | 2.12 |
| Alessi 53 | -2.66 | 255 | 365 | 19.3 |
| Juchert 12 | -2.40 | 125 | 250 | 13.5 |
| Teutsch 144 | -2.70 | 745 | 890 | 12.6 |

Comparing the real ages of these clusters with their dynamical relaxation times, we found that the ages are 23, 26, 22 and 63 times of the relaxation ones. Thus we can conclude that these clusters are dynamically relaxed.

## 7. Conclusions

According our analysis for testing and estimating the fundamental parameters of some newly discovered stellar groups of Kronberger et al. (2006) using *2MASS* photometry, we ensured that (Teutsch 144, Alessi 53, Riddle 4 and Juchrt 12) are really open star clusters, which have *IMFs* around the Salpeter's (1955) value. On the other hand, the real ages of these clusters are found to be more than 20 times (at least) of their relaxation ones, which infer that these clusters are dynamically relaxed. The main parameters of these clusters are listed in Tables 2 and 3.